# Experimental observation of effective gravity and two-time physics in ferrofluid-based hyperbolic metamaterials

V. N. Smolyaninova [a], J. Cartelli [a], B. Augstein [a], S. Spickard [a], M. S. Devadas [a],

I. I.Smolyaninov [b]

[a] *Department of Physics Astronomy and Geosciences, Towson University,*

*8000 York Rd., Towson, MD 21252, USA*

[b] *Department of Electrical and Computer Engineering, University of Maryland, College Park, MD 20742, USA ; smoly@umd.edu*

**Abstract.** Recently it was predicted that extraordinary light waves in hyperbolic metamaterials may exhibit two-time physics behavior. We report experimental observation of this effect via investigation of gravity-like nonlinear optics of iron/cobalt-based ferrofluid hyperbolic metamaterials. In addition to conventional temporal coordinate, the spatial coordinate oriented along the optical axis of the metamaterial also exhibits timelike character, which leads to very unusual two-time physics behavior in these systems on small scales.

**Keywords:** hyperbolic metamaterial, nonlinear optics, ferrofluid

## 1. Introduction

Electromagnetic metamaterials have enabled new ways to control and manipulate electromagnetic waves. One of the more unusual applications of metamaterials was the recent theoretical proposal to construct a physical system, which would exhibit two-time (2T) physics behavior on small scales [1]. Theoretical investigation of the 2T space-time models had been pioneered by Dirac [2] and Sakharov [3]. More recent examples may be found in [4, 5]. However, to our knowledge, a real experimental system which would exhibit such a 2T behavior was never demonstrated in the experiment.

The theoretical proposal in [1] considered propagation of monochromatic extraordinary waves in a hyperbolic metamaterial [6], which is described by anisotropic dielectric tensor having the diagonal components $\varepsilon_{xx} = \varepsilon_{yy} = \varepsilon_1 > 0$ and $\varepsilon_{zz} = \varepsilon_2 < 0$. The wave equation in such a metamaterial may be written as

$$-\frac{\partial^2 \varphi_\omega}{\varepsilon_1 \partial z^2} + \frac{1}{(-\varepsilon_2)}\left(\frac{\partial^2 \varphi_\omega}{\partial x^2} + \frac{\partial^2 \varphi_\omega}{\partial y^2}\right) = \frac{\omega_0^2}{c^2}\varphi_\omega = \frac{m^{*2}c^2}{\hbar^2}\varphi_\omega \quad , \qquad (1)$$

where $\varphi_\omega = E_z$ is the photon electric field component parallel to the optical axis $z$ of the hyperbolic metamaterial. This wave equation coincides with the Klein-Gordon equation for a massive field $\varphi_\omega$ (with an effective mass $m^*$) in a 3D Minkowski spacetime in which one of the spatial coordinates $z = \tau$ behaves as a timelike variable. The metric coefficients $g_{ik}$ of this flat 2+1 dimensional Minkowski spacetime may be defined as [1,7]:

$$g_{00} = -\varepsilon_1 \qquad \text{and} \qquad g_{11} = g_{22} = -\varepsilon_2 \qquad (2)$$

As demonstrated in [8], nonlinear optical Kerr effect "bends" this 2+1 Minkowski spacetime, resulting in effective gravitational force between extraordinary photons. It was also predicted that for the effective gravitational constant inside the metamaterial to be positive, negative self-defocusing Kerr medium must be used as a dielectric host of the metamaterial [8]. Experimental observation of the effective gravity in such a system should enable observation of the emergence of the gravitational arrow of time along the $z$ direction, which is predicted to occur even at the classical gravity level [9]. Such an observation would further extend experiments on the emergence of the arrow of time in planar hyperbolic metamaterials reported in [7].

On the other hand, the model of 2+1 metamaterial Minkowski spacetime described above may itself evolve in our own physical time, which means that a theoretically proposed metamaterial-based 2T physics system [1] will be realized and observed in the experiment. In this paper we report an experimental study of gravity-like nonlinear optical effects in a three-dimensional self-assembled hyperbolic metamaterials based on iron-cobalt ferrofluid subjected to external magnetic field. Magnetic interaction of the iron-cobalt nanoparticles in the ferrofluid is rather weak, so in the absence of external magnetic field the nanoparticles are randomly distributed within the fluid, as illustrated in Fig. 1a. On the other hand, application of a modest external magnetic field (of the order of 300 Gauss) leads to formation of nanocolumns aligned along the external field (see Fig. 1b), thus leading to formation of a self-assembled 3D metamaterial, which exhibits hyperbolic behaviour in the long wavelength infrared (LWIR) range [10]. The ferrofluid nanocolumns were imaged in Fig.1f of [10]. The typical distance between the nanocolumns was measured to be ~0.3 $\mu$m, which is much smaller than the 10.6 $\mu$m wavelength of $CO_2$ laser used in our experiments, and which justifies the metamaterial description of the ferrofluid in this spectral range. Even though the arrangement of nanocolumns observed in the experiment was not strictly periodic, the hyperbolic metamaterial description is known to hold for such structures [6], which was confirmed in [10]. While the Fe-Co nanoparticles are magnetic, the AC magnetic response of the self-assembled chains of these nanoparticles inside the ferrofluid at optical frequencies is negligible. Even the magnetic response of individual nanoparticles in the chains is known to cut off above ~10 MHz [11]. Moreover, kerosene used in our experiments as a carrier fluid is non-magnetic. Therefore, the theoretical description of such a metamaterial from [1,8] does not need to be revised. Even though technically, the time-reversal symmetry in this material system is broken, and the studied hyperbolic metamaterial should become nonreciprocal, in reality these effects are very weak. For example, reversal of the external magnetic field and the individual magnetic moments of the nanoparticles under a time-reversal operation leads to formation of exactly the same "nanowire" hyperbolic metamaterial.

Similar to other common fluids, kerosene (used as a host fluid for the cobalt nanoparticles) exhibits negative self-defocusing Kerr effect in the strong optical field of a $CO_2$ laser operating at the 10.6 μm wavelength. Therefore, in agreement with theoretical predictions [8], the fabricated ferrofluid indeed exhibited pronounced gravity-like self-focusing effects, leading to emergence of the gravitational arrow of time [9] oriented along the optical axis $z$ of the metamaterial. On the other hand, absorption of the LWIR light in the metamaterial leads to gradual heating of the system, thus leading to irreversible evolution of the system as a function of conventional physical time. A detailed account of these experiments, which realize the theoretically proposed 2T physics scheme [1] is reported below.

## 2. Materials and Methods

Schematic diagram of our experimental setup is shown in Fig. 1c. A sample of ferrofluid is contained between two NaCl windows separated by a thin (100 μm) spacer within an optical cuvette and secured vertically to the optical table. A horizontally polarized $CO_2$ laser beam is then passed through the sample and a LWIR camera (FLIR Systems) is used to image the beam shape after its passage through the ferrofluid (it was verified that the original beam shape of the laser may be well described by a simple Gaussian profile - see inset in Fig. 1c. To see the effect of a DC magnetic field on the beam shape, which is associated with formation of a self-assembled hyperbolic metamaterial inside the ferrofluid, a large permanent magnet is placed near the cuvette (above or to the side), producing a magnetic field aligned almost vertically or almost horizontally (but not parallel to the cuvette). As a result, the laser light passing through the metamaterial becomes either predominantly ordinary (E field of the light field oriented perpendicular to the optical axis of the metamaterial), or predominantly extraordinary (E field having non-zero component parallel to the optical axis). Temporal evolution of the beam shape was studied as a function of magnetic field direction and laser intensity.

The iron-cobalt ferrofluid used in these experiments was similar to the cobalt ferrofluid used in [10], since the same carrier fluid (kerosene) was used in both cases and the optical properties of cobalt and iron are similar in the LWIR range [12]. The ferrofluid fabrication will be described elsewhere. Detailed microscopic and spectroscopic characterization of the structural and frequency-dependent dielectric properties of the cobalt-based ferrofluid may be found in [10], where we have reported imaging of the nanocolumns formed in the ferrofluid under the influence of external magnetic field, and the detailed study of the anisotropic dielectric tensor of the ferrofluid. Within the 5-8 % volume fraction of the iron-cobalt nanoparticles in the ferrofluid (which is typical for our experiments) the self-assembled nanocolumn structure exhibits hyperbolic properties in the LWIR spectral range. According to the theoretical predictions [13-16], these hyperbolic properties should give rise to quite counter-intuitive nonlinear optical properties of the ferrofluid. These reports demonstrated existence of spatial solitons in an array of nanowires embedded in a Kerr medium. However, an interesting counter-intuitive feature of these solitons is that they occur only if a self-defocusing negative Kerr medium is used as a dielectric host. It was demonstrated in a later paper [8] that behavior of spatial solitons in nonlinear hyperbolic metamaterials finds natural explanation in terms of analogue gravity. Indeed, nonlinear optical Kerr effect in a hyperbolic metamaterial "bends" the effective optical Minkowski spacetime described by Eqs. (1,2), resulting in effective gravitational force between extraordinary photons. In order for the effective gravitational constant to be positive, negative self-defocusing Kerr medium must be used as a host [8].

When the nonlinear optical effects become important, they are described in terms of various order nonlinear susceptibilities $\chi^{(n)}$ of the metamaterial. Taking into account these nonlinear terms, the dielectric tensor of the metamaterial (which defines its effective metric) may be written as

$$\varepsilon_{ij} = \chi_{ij}^{(1)} + \chi_{ijl}^{(2)} E_l + \chi_{ijlm}^{(3)} E_l E_m + ... \qquad (3)$$

Similar to general relativity, Eq. (3) provides coupling between photons and the effective metric of the metamaterial "spacetime". In a central symmetric material the second order susceptibility equals zero, while the third order terms may provide correct coupling between the effective metric and the energy-momentum tensor in the weak field limit. These terms are associated with the optical Kerr effect [8]. Indeed, in the weak gravitational field limit the Einstein equation is reduced to

$$R_{00} = \frac{1}{c^2} \Delta \phi = \frac{1}{2} \Delta g_{00} = \frac{8\pi\gamma}{c^4} T_{00} \quad , \qquad (4)$$

where $\phi$ is the gravitational potential [17]. Since in the effective optical spacetime the timelike coordinate is identified with the optical axis direction $z$, and $g_{00}$ is identified with $-\varepsilon_1$, comparison of Eqs. (3) and (4) performed in [8] indicated that the effective gravitational constant $\gamma^*$ may be identified as

$$\gamma^* \approx \frac{\chi^{(3)} c^2 \omega^2 \varepsilon_2}{4} = \frac{\pi^2 c^4 \chi^{(3)} \varepsilon_2}{\lambda_0^2} \quad , \qquad (5)$$

where $\lambda_0$ is the free space wavelength (see Eq. (16) from [8]). Since $\varepsilon_{zz} = \varepsilon_2 < 0$, in the nanowire array metamaterial, the sign of $\chi^{(3)}$ must be negative for the effective gravity to be attractive. Extraordinary light rays in such a medium will behave as 2+1 dimensional world lines of gravitating (and therefore self-gravitating) bodies, and these rays may exhibit self-focusing behavior, which is somewhat analogous to gravity [8]. Because of the large and negative thermo-optic coefficient inherent to most liquids, heating produced by partial absorption of the propagating laser beam translates into a significant decrease of the refractive index at higher light intensity. For example, reported thermo-optics coefficient of kerosene reaches $\Delta n/\Delta T= -1.6 \times 10^{-4} K^{-1}$ [18]. Therefore, ferrofluid-based self-assembled hyperbolic metamaterials appear to be an ideal material choice to demonstrate these effects in the experiment.

The electromagnetic properties of the metamaterial used in our experiments are defined by its dielectric permittivity components [10]:

$$\varepsilon_2 = \alpha\varepsilon_m + (1-\alpha)\varepsilon_d \qquad \text{and} \qquad \varepsilon_1 = \frac{2\alpha\varepsilon_m\varepsilon_d + (1-\alpha)\varepsilon_d(\varepsilon_d + \varepsilon_m)}{(1-\alpha)(\varepsilon_d + \varepsilon_m) + 2\alpha\varepsilon_d} \qquad (6)$$

where $\alpha$ and $\varepsilon_m$ are the volume fraction and the dielectric permittivity of nanoparticles, and $\varepsilon_d$ is the dielectric permittivity of kerosene. They were characterized extensively in [10]. It was demonstrated that at 10.6 μm the $\varepsilon_{zz}=\varepsilon_2$ component of the metamaterial dielectric tensor equals approximately -70, and its magnitude is primarily defined by the volume fraction of magnetic nanoparticles in the ferrofluid (since $\varepsilon_m >> \varepsilon_d$, see Eq. (6)). On the other hand, the $\varepsilon_{xx}=\varepsilon_{yy}=\varepsilon_1$ components of the metamaterial tensor are approximately equal to the dielectric permittivity of kerosene $\varepsilon_d$=2.3 (see Eq.(6) and Fig.3d from [10]). Since the boiling point of kerosene is 175ºC, and it was verified that kerosene was not boiling at the $CO_2$ laser power used in our experiments, the maximum change of the kerosene refractive index cannot exceed $\Delta n$=0.024. While really huge by the nonlinear optics standards, even such a large $\Delta n$ cannot change the hyperbolic character of the metamaterial.

On the other hand, according to Eq.(5), very large magnitudes of $\chi^{(3)}$ and $\varepsilon_2$ make the effective gravity and the self-focusing effects very pronounced in the studied metamaterial. For example, in a conventional isotropic Kerr medium self-focusing occurs if Gaussian beam power $P$ is greater than the critical power [19]

$$P_{cr} = \frac{1.9 n_0}{4\pi n_2} \left( \frac{\lambda_0}{n_0} \right)^2 , \tag{7}$$

where $n_0$ and $n_2$ are the linear and nonlinear refractive indices of the medium and $\lambda_0/n_0$ is proportional to the diffraction-limited radius of the self-focused beam. The corresponding result in a nonlinear hyperbolic medium has been derived in [8] using the effective gravity analogy:

$$P_{cr} \propto \frac{\varepsilon_1^0}{\left( -\chi^{(3)} \right)} \rho^2 , \tag{8}$$

where $\rho$ is the radius of self-focused beam which is no longer diffraction-limited in a hyperbolic metamaterial (see Eq.(22) from [8]). It must be assumed to be equal to the structural parameter of the metamaterial. As a result, self-focusing in a hyperbolic medium occurs at much lower power levels compared to the conventional Kerr media. Indeed, thermo-optical nonlinear refractive index $n_2 = 7.4 \times 10^{-8}$ cm$^2$/W of a somewhat similar gold nanoparticle suspension in castor oil has been measured in [20]. Together with the structural parameter $\rho=0.3$ μm measured in [10] for our ferrofluid-based metamaterial, this leads to an estimate of the critical self-focusing power of the order of $P_{cr} \sim 3$mW. The nonlinear focusing distance $Z_{NL}$ for a longitudinal wave vector component $k$ may be also estimated as usual:

$$Z_{NL} = n_0/k n_2 I , \tag{9}$$

where $I$ is the $CO_2$ laser light intensity. Since the $k$ vector in a hyperbolic metamaterial is not diffraction-limited, $k \sim 1/\rho$ produces $Z_{NL} \sim 20$ μm. These estimates of the critical power and the nonlinear focusing distance are consistent with our experimental results described below.

## 3. Results.

Experimentally measured temporal dependencies of the $CO_2$ laser beam shape for the ordinary (Fig.2a, Video 1) and extraordinary (Fig.2a, Video 2) light passing through the ferrofluid subjected to external DC magnetic field indeed agree well with the theoretical picture predicted in [8, 13-16]. These videos were recorded at the same 160 mW incident laser power. While the ordinary light beam shape hardly experiences any changes over time, the extraordinary beam separates into multiple filaments which exhibit fast dynamical behaviour on the millisecond time scale. It has been verified that this filamentation disappears at lower light power (compare Fig.2a Videos 4 and 5 taken at 160 mW and 22 mW incident $CO_2$ laser power, respectively). The dynamical filaments were also not observed in the absence of external DC magnetic field when the ferrofluid was non-structured and isotropic (see Fig.2a Video 3). The observed filamentation of the extraordinary beam is also illustrated in Fig.2a, which shows a representative image frame from Video 2, and its cross section. The absence of filamentation in the cases of ordinary light propagating through either isotropic or anisotropic ferrofluid is a natural consequence of strong negative self-defocusing Kerr effect in kerosene. According to [8, 13-16], self-focusing only happens in the case of extraordinary light propagation.

The beam filamentation of the extraordinary light may be revealed more clearly in the differential images, in which the average Gaussian profile of the laser beam is subtracted from the currently observed beam shape. An example of such a differential image is presented in Fig. 2b. It reveals just the filaments, which are present in the extraordinary beam at a given moment of time. It is interesting to note that FFT analysis of Fig. 2b, which is shown in Fig. 2c, reveals somewhat perturbed hexagonal symmetry in the spatial distribution of the filaments. Such a hexagonal symmetry in filament distribution is quite common in other nonlinear optical systems exhibiting self-focusing behaviour [21]. The highly symmetric spatial pattern of filaments is better revealed after high pass filtering of Fig. 2b, which is performed in Fig. 2d.

As illustrated in Fig. 3, these results demonstrate emergence of the "gravitational arrow of time" in the self-assembled hyperbolic metamaterial, which is directed along the time-like spatial z coordinate (aligned with the optical axis of the metamaterial). Indeed, as demonstrated in [9], in the presence of gravity a simple Newtonian system of dust particles develops progressively more complex structures, so that "the growth-of-complexity arrow" always points away from the unique past. It is very natural to identify the arrow of time with the direction in which the complexity of the structures grows. As emphasized in [9], complexity is a prerequisite for storage of information in local subsystems and therefore the formation of records. Under the action of gravity typical systems break up into disjoint subsystems which get more and more isolated, and one can associate dynamically generated local information with them. This time evolution is depicted in Fig. 3a for an example of a planetary system forming from a structureless dust cloud. The observed changes in shape of the extraordinary light beam as a function of $z$ coordinate (depicted in Fig. 3b) exhibit virtually similar behaviour. The original structureless Gaussian beam incident onto the cuvette separates into multiple filaments under the action of effective gravity acting in the effective 2+1 dimensional optical spacetime, in which the role of time is played by the spatial $z$ coordinate.

On the other hand, as evident from Video 2, the spatial pattern of filaments exhibits fast variations as a function of the conventional physical time on the millisecond time scale. For example, four image frames taken from such a video (see Fig. 4a) show gradual motion of two filaments towards each other, followed by merging of these filaments. In addition, absorption of the $CO_2$ laser light in the metamaterial causes gradual heating of the system, thus leading to irreversible evolution of the system as a function of the conventional physical time. Fig. 4b shows an example of such an entropic irreversible evolution of the extraordinary beam following the opening of the $CO_2$ laser shutter. The average temperature of the field of view measured by the LWIR camera increases from frame to frame. Note that mutual attraction of individual filaments may also be seen in this set of images.

## 4. Discussion

Our experimental observations clearly reveal the 2T character of the ferrofluid-based hyperbolic metamaterial system on a small scale. As theoretically predicted in [1], propagation of extraordinary light in this system is described by two time-like coordinates. In addition to the conventional physical time, the spatial direction aligned with the optical axis $z$ of the metamaterial also has a time-like character, which is revealed by the Minkowski-like 2+1 dimensional metric of the optical space described by Eqs. (1,2) and by our experimental observations of the emergence of the gravitational arrow of time aligned along the $z$ direction. Thus, the presented experimental system appears to demonstrate quite unique dynamical properties, which do not have any analogs among other natural or artificial materials. It should be useful in our attempts to understand the nature of time, causality and the meaning and interplay of various "arrows of time" (thermodynamic, gravitational, cosmological, etc.) introduced in different fields of physics. For example, we should point out that the concept of ferrofluid-based "metamaterial multiverse" introduced in [22] closely resembles the Sakharov's concept of multiverse described in [3] as Minkowski-like regions of space-time continuum inside a larger purely spatial multi-dimensional

"Parmenidian" (or "Euclidean") region. Sakharov's justification for considering the physics of 2T (or multi-time) space-time regions comes from the suggestion that the observed universe and an infinite number of other universes arose as a result of quantum transitions inside the multi-dimensional Euclidean space, accompanied by a change in the signature of the metric. Sakharov demonstrated that the issues related to causality in such multi-time models may be resolved by compactification of the additional temporal dimensions. We note that the 2T physics of ferrofluid-based hyperbolic metamaterials considered above is consistent with this picture because of high losses, and therefore very short propagation length of the extraordinary photons inside the metamaterial. These high losses also affect reflections off the metamaterial boundaries, which has been studied in detail in [23]. It was demonstrated that the extraordinary field is strongly enhanced near the hyperbolic media boundaries (see for example Fig.3 from [23]). In combination with high metamaterial losses this field enhancement leads to strong suppression of reflections from the boundaries.

We also note that practical applications of the 2T systems may be also quite interesting. For example, they may be applied in novel optical hyper-computing schemes [24], which map a computation performed during a given period of time onto a much faster computation performed using a given spatial volume of a hyperbolic metamaterial. Such hyper-computing schemes were described in detail in [24]. They may be useful in time-sensitive applications. Since practical applications of such systems are limited by propagation losses, we should note that losses may be reduced if a better plasmonic metal is used in nanoparticles. However, such good plasmonic metals as gold and silver are non-magnetic. While there were attempts at creating composite nanoparticles made of gold in combination with some magnetic metals, they are not highly successful yet.

In conclusion, the introduction of transformation optics [25] was a hugely important novel advance in optics, which has led to many breakthrough developments (such as optical invisibility cloaking, super-resolution microscopy, etc.). The central concepts of transformation optics are the concepts of "optical space" and "optical metrics", which differ from the metric of physical space. The concept of effective "optical spacetime" in hyperbolic metamaterials [6, 26] is one of the most important tools developed in transformation optics to describe often counterintuitive optics of hyperbolic metamaterials. The current work demonstrates successful extension of this concept into the nonlinear optical domain. As demonstrated by our results, effective gravity appears to be the most natural mathematical language to describe perturbations of the effective optical spacetime in nonlinear hyperbolic metamaterials.

**Figure Captions**

**Figure 1.** (a) In the absence of external magnetic field, cobalt nanoparticles are randomly distributed within the ferrofluid, and their magnetic moments (which are shown by red arrows) have no preferred spatial orientation. (b) Application of external magnetic field leads to formation of nanocolumns (made of nanoparticles) which are aligned along the field direction. (c) Schematic diagram of the experimental geometry. A LWIR camera (FLIR Systems) is used to study $CO_2$ laser beam propagation through the ferrofluid subjected to external DC magnetic field. The inset shows the measured beam shape in the absence of the ferrofluid sample. Two orientations of the external magnetic field $B$ used in our experiments are shown by green arrows. The red arrow show laser light polarization. Note that z-axis is assumed to always point in the direction of the external DC magnetic field.

**Figure 2**. (a) Comparison of experimentally measured temporal dependences of the $CO_2$ laser beam shape for the ordinary light (see Video 1) and extraordinary light (see Video 2) passing through the ferrofluid subjected to external DC magnetic field (measured at 160 mW laser power). Video 3 shows a similar temporal signal dependence for the light passing through the isotropic ferrofluid, which is not subjected to any external DC magnetic field. Filamentation of the extraordinary beam is clearly visible in a single 2.6 mm x 2.6 mm frame taken from the Video 2, and its cross section (see Video 4). The cross section of the ordinary beam (taken from Video 1) measured at the same laser power is shown for comparison. Video 5 shows experimentally measured temporal dependence of the CO2 laser beam profile for the extraordinary light polarization recorded at 22 mW incident CO2 laser power. At this lower power the beam filamentation virtually disappears, so that the beam shape may be characterized as slightly perturbed Gaussian profile. (Video 1, MP4, 2.8 MB; Video 2, MP4, 5.0 MB; Video 3 MP4, 3.4 MB; Video 4, MP4, 1.7 MB; Video 5, MP4, 1.7 MB) (b) The beam

filamentation of the extraordinary light is revealed more clearly in the differential image, in which the average Gaussian profile of the laser beam is subtracted from the currently observed beam shape. (c) FFT analysis of image (b) reveals somewhat perturbed hexagonal symmetry in the spatial distribution of the filaments. The corresponding Fourier components are highlighted by white dots. (d) The highly symmetric spatial pattern of filaments is better revealed after high pass filtering of image (b).

**Figure 3.** (a) In the presence of gravity a simple Newtonian system of dust particles develops progressively more complex structures, so that "the growth-of-complexity arrow" always points away from the unique past. This time evolution is depicted for an example of a planetary system forming from a structureless dust cloud. (b) The observed changes in shape of the extraordinary light beam as a function of z coordinate exhibit virtually similar behavior. The original structureless Gaussian beam incident onto the cuvette separates into multiple filaments under the action of effective gravity acting in the effective 2+1 dimensional Minkowski spacetime, in which the role of time is played by the spatial z coordinate.

**Figure 4.** (a) Four 2.6 mm x 2.6 mm image frames taken from a recorded video of the extraordinary beam evolution as a functional of conventional physical time (the frame number is shown in the corner of each image). These frames show gradual motion of two filaments (indicated by arrows) towards each other, followed by merging of these filaments. (b) An example of an entropic irreversible temporal evolution of the extraordinary beam following the opening of the $CO_2$ laser shutter. The average temperature of the field of view measured by the LWIR camera increases from frame to frame. The dimensions of all the frames are 2.6 mm x 2.6 mm. All the time frames are recorded in the same physical location. They correspond to the optical field distribution

at the output face of the optical cuvette, which is located at 100 μm distance from the input face of the cuvette.

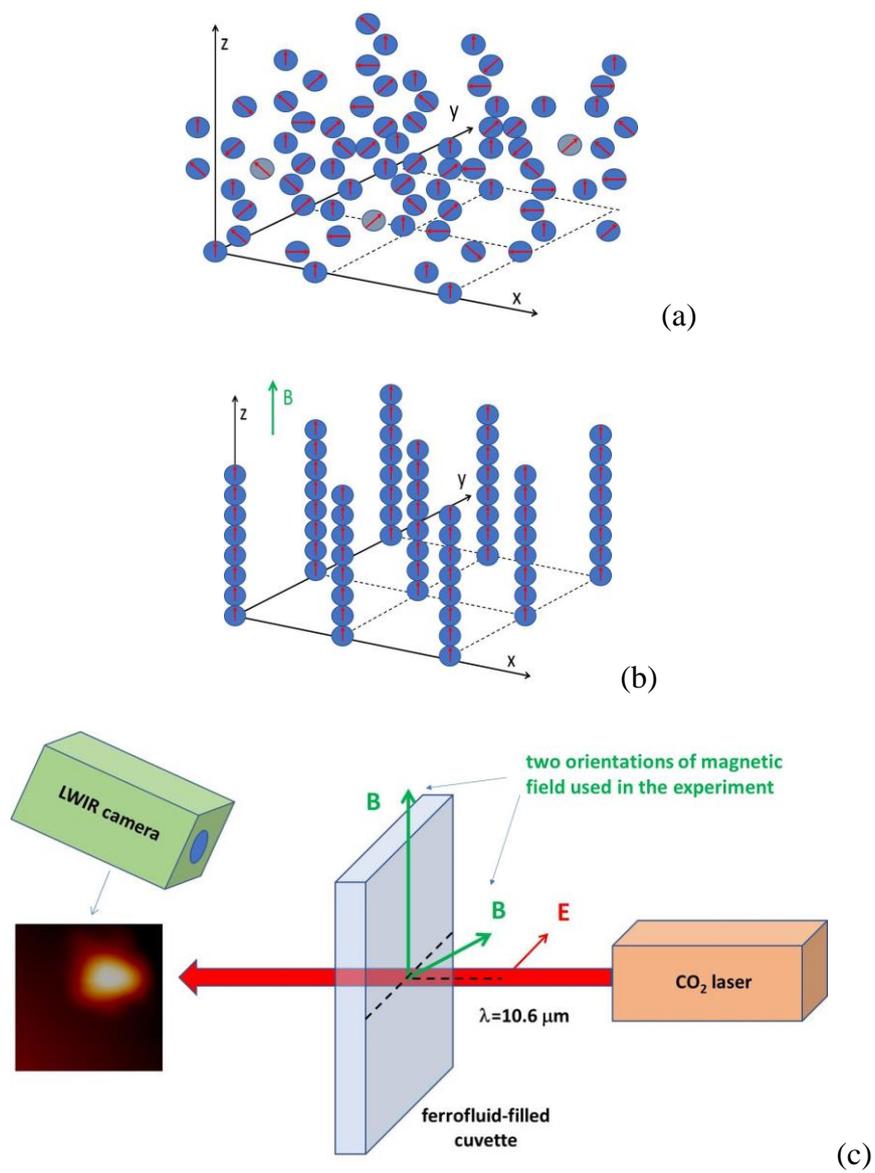

Fig. 1

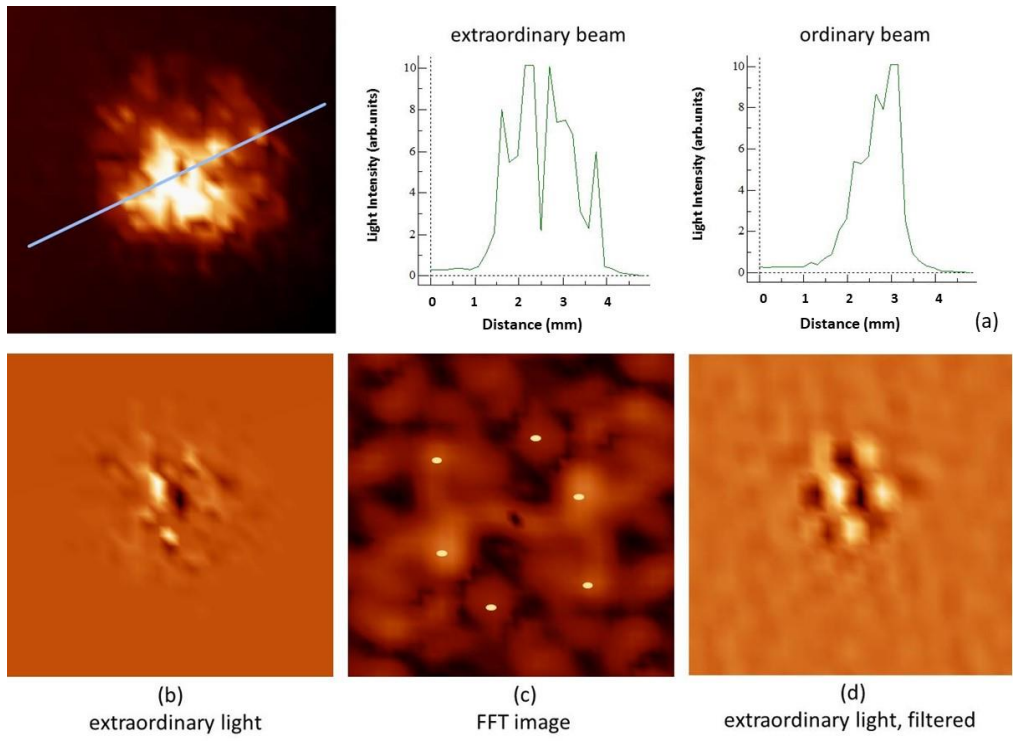

(a)

(b)
extraordinary light

(c)
FFT image

(d)
extraordinary light, filtered

Fig. 2

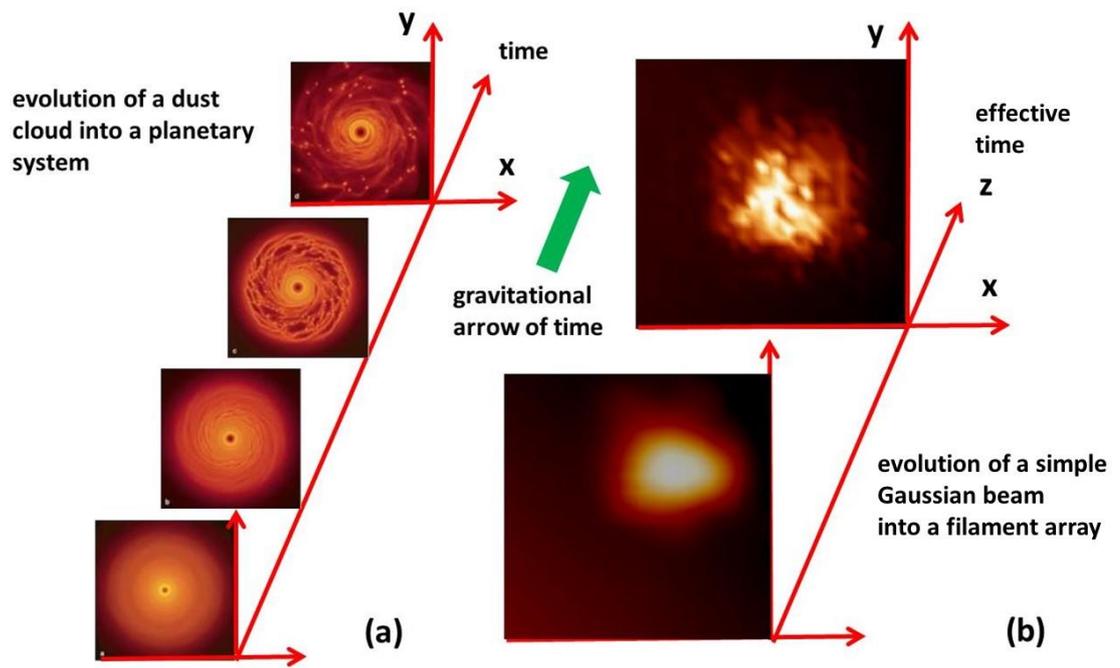

**evolution of a dust cloud into a planetary system**

time

y

x

gravitational arrow of time

**(a)**

y

effective time

z

x

**evolution of a simple Gaussian beam into a filament array**

**(b)**

Fig. 3

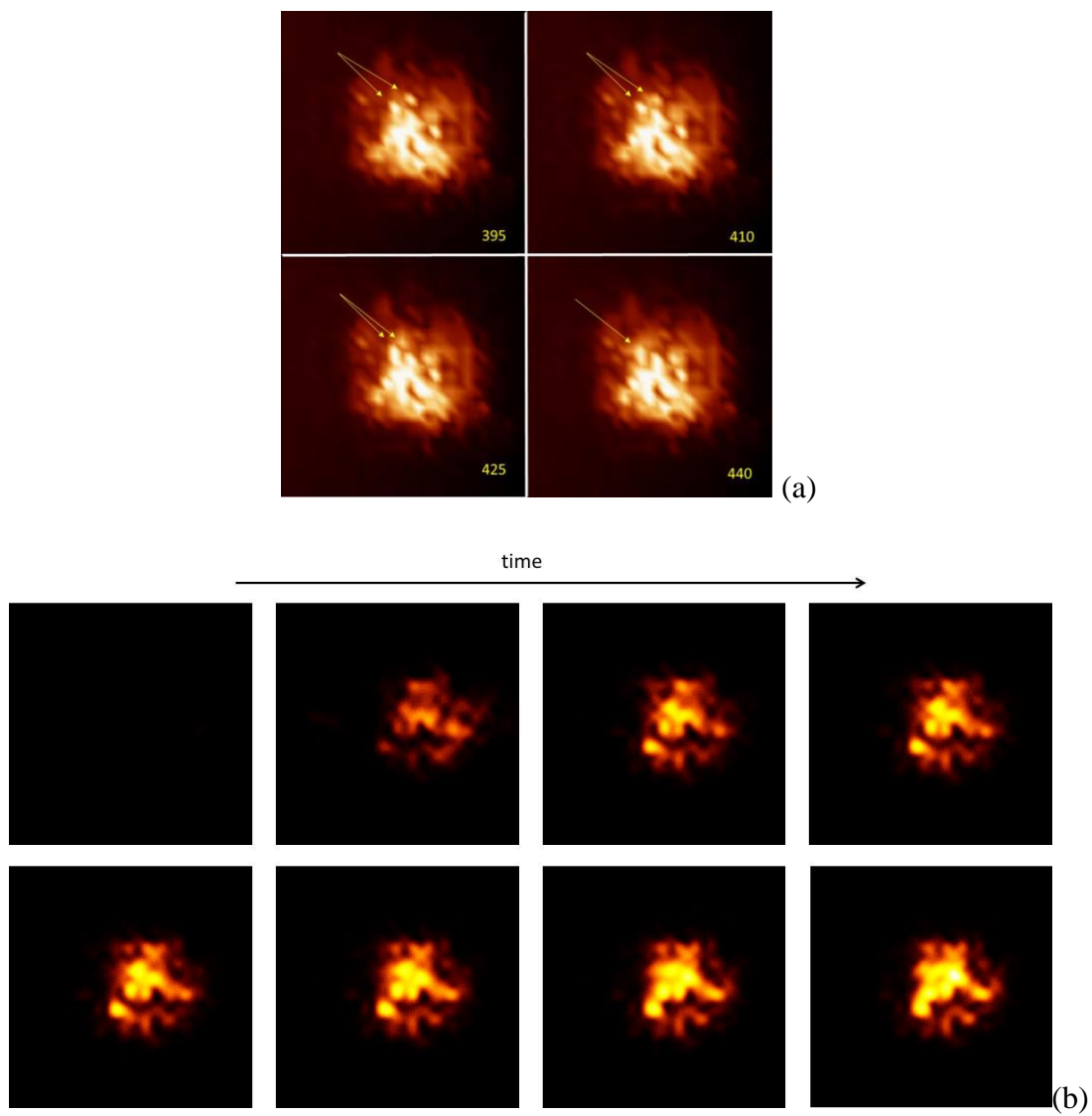

Fig. 4